\documentclass[11pt]{article}
\usepackage[utf8]{inputenc}
\usepackage{amsmath}
\usepackage{graphicx}
\usepackage{subcaption}
\usepackage{authblk}
\usepackage{xcolor}
\usepackage{mathptmx}
\usepackage[page,toc,titletoc,title]{appendix}
\usepackage{comment}
\usepackage[overload]{empheq}
\usepackage{cases}
\usepackage{lipsum}
\usepackage{amsfonts}
\usepackage{hyperref}
\usepackage{float}
\usepackage{atbegshi}

\usepackage[a4paper, includehead = True, includefoot = True]{geometry}
\newcommand{\dd}{\mathrm{d}}
\newcommand{\meanu}{\bar{u}}
\newcommand{\SSiw}{\mathrm{S}_{iw}}
\newcommand{\SSi}{\mathrm{S}_i}
\newcommand{\SSS}{\mathrm{S}}
\newcommand{\SSw}{\mathrm{S}_w}

\newcommand{\TKE}{\mathrm{TKE}}

\title{Steady-state supersaturation distributions for clouds under turbulent forcing}

\begin{document}

\author[1]{Manuel Santos Guti\'errez \thanks{Corresponding author. Email: \texttt{manuel.santos-gutierrez@weizmann.ac.il}}}
\author[2]{Kalli Furtado\thanks{Email: \texttt{kalli\_furtado@nea.gov.sg}}} 
\affil[1]{Department of Earth and Planetary Sciences, Weizmann Institute of Science, Rehovot, Israel}
\affil[2]{Centre for Climate Research Singapore, Meteorological Service Singapore, Singapore}

\maketitle

\begin{abstract}

The supersaturation equation for a vertically moving adiabatic cloud parcel is analysed. The effects of turbulent updrafts are incorporated in the shape of a stochastic Lagrangian model, with spatial and time correlations expressed in terms of turbulent kinetic energy. Using the Fokker-Planck equation, the steady-state probability distributions of supersaturation are analytically computed for a number of approximations involving the timescale separation between updraft fluctuations and phase-relaxation, multiplicative noise simplification and droplet or ice particle size fluctuations. While the analytical results are presented in general for single-phase clouds, the calculated distributions are used to compute mixed-phase cloud properties--- mixed fraction and mean liquid water content in an initially icy cloud--- and are argued to be useful for generalising and constructing new parametrisation schemes.

\end{abstract}

\tableofcontents

\section{Introduction}

Water vapour supersaturation is a key thermodynamic parameter in the formation and development of warm and cold clouds. The activation of cloud condensation nuclei (CCN), the spectrum of droplet sizes and, therefore, precipitation heavily depend on the supersaturation budget of an evolving cloud. By extension, the study of supersaturation is fundamental to determine the radiative properties of cloud fields which constitute a major climate feedback \cite{khain_pinsky_2018}, yet a major source of uncertainty in climate projections \cite{zelinka2020}.

Due to its microphysical character, supersaturation cannot be explicitly resolved by global circulation models (GCMs) and, hence, it has to be prescribed using adequate parametrisation techniques; see the seminal work in \cite{arakawa1974}. One way of doing this is by predicting the probability of encountering the cloud parcel at a certain relative humidity level in terms of the background dynamical and large-scale information. Taking the GCM perspective, the idea is to infer the subgrid cloud properties out of the prognosed variables: vertical velocity, turbulent kinetic energy, temperature, humidity and pressure, which altogether depict the relative humidity configuration of an homogeneous cloud parcel. However, localised temperature gradients or turbulence at the smaller scales can create inhomogeneous fluctuations which alter the spread of supersaturation values, hence changing the microphysical properties of the subgrid cloud parcel. This is particularly relevant in mixed-phase clouds, where the turbulent dynamics influence the activation of long-lived supercooled water in icy clouds \cite{Korolev2008,Hill2014, Field2014}. 

Different updraft profiles--- ranging from steady to turbulent--- yield different supersaturation distributions which determine key features like mean liquid water content, mixed-phase cloud fraction or the activation fraction of CCN. It is crucial, then, to determine not only the mean value of supersaturation at given location, but also its variance or even higher moments. To this end, the equations for supersaturation or condensational growth are coupled with a suitable stochastic forcing law that captures the effects of turbulence \cite{bartlett1972,shaw_2003}. Hence, the classical deterministic models become stochastic differential equations (SDEs), which have been widely used in the cloud physics community \cite{shaw_2003,khain_pinsky_2018}. In fact, this stochastic physics framework has been used in the elaboration of analytically tractable parametrisation schemes for mixed-phase clouds \cite{Field2014}, the study of droplet growth by condensation \cite{bartlett1972,sardina_2015} and the determination of steady-state warm cloud properties \cite{siewert_bec_krstulovic_2017}. Departing from the work of \cite{Field2014}, we extend their analytical predictions of supersaturation distribution to a wider range of contexts, involving the equations of turbulent updrafts, supersaturation and diffusional growth. 

This paper is structured as follows. In Section~\ref{sec:2}, the Squires equation for the evolution of supersaturation is revisited from first principles and, secondly, a stochastic equation for turbulent updrafts is presented, in the lines of the theory of stochastic Lagrangian turbulent models; see, e.g., \cite{rodean1996}. In Section~\ref{sec:3}, the quasi-steady equation--- which assumes a constant mean droplet/particle radius--- is analysed in a number of approximations providing formulas for the probability distribution of supersaturation. In Section \ref{sec:diffusional growth}, fluctuations in droplet size are allowed and their net effects on supersaturation distribution are investigated in analytical terms. A total of five different supersaturation distributions are obtained on analytical grounds. In Section \ref{sec:parametrisation}, the relevance of the five obtained probability density functions is discussed and compared in the context of mixed-phase clouds. Finally, a discussion over the results is done in Section \ref{sec:discussion}. To supplement the information in the main text, appendices are included to discuss some technical topics related to the analysis of stochastic differential equations.

\section{The supersaturation equation}\label{sec:2}
\subsection{The Squires equation}

We consider a vertically moving cloud parcel containing a monodisperse family of liquid water droplets or ice particles that are spatially uniformly distributed. Furthermore, it is assumed that droplet/particle number remains constant in time. The evolution of supersaturation in the cloud is, essentially, determined by the sources and sinks of relative humidity due to the adiabatic cooling of the parcel in ascent, and condensation of water vapour onto the existing droplets' or particles' surface \cite{rogersyau}. However, the exact relation between the rate of change of supersaturation and its sinks and sources is obtained by taking its derivative with respect to to time. We recall that supersaturation, for either ice or water--- is defined as:
\begin{equation}\label{eq:superaturation}
	\SSS = \frac{e - E}{E},
\end{equation}
where $e$ is the water vapour pressure and $E$ the same although at saturation over a flat surface of liquid water or ice. We shall not specify now whether we are dealing with water or ice supersaturation because the stochastic analysis will be done independently. It is noted, however, that the calculations immediately below can be done for single or mixed-phase clouds; see \cite{Korolev2003}. 

Taking the time derivative of $\SSS$ and employing the mass conservation, temperature, Clausius-Clayperon and the quasi hydrostatic approximation, P.~Squires derived in 1952 an equation to describe the evolution of the supersaturation budget \cite{squires_1952}--- see also the Appendix in \cite{Korolev2003}:
\begin{equation}\label{eq:supersaturation equation}
\frac{1}{1 + \SSS} \dd \SSS = au - b\frac{\dd q}{\dd t},
\end{equation}
where $u$ and $q$ are the vertical velocity and liquid-water or ice mixing ratio, respectively. The evolution of $\SSS$ obeys an equation with a nonlinear term stemming from the time and temperature dependence of the equilibrium water vapour pressure $E$ in Eq.~\eqref{eq:superaturation}. The equation \eqref{eq:supersaturation equation} reveals that only to leading order in small values of $\SSS$ do we obtain a linear dependence on vertical velocity $u$ and water vapour condensation/deposition, $\dd q/\dd t$. Such is the case of warm clouds, where supersaturation levels do not typically exceed $2\%$ \cite{prabha_2011}. However, mixed-phase conditions arise precisely when ice supersaturation fluctuates fully icy clouds spreading beyond small values \cite{Field2014}, making the nonlinearity in Eq.~\eqref{eq:supersaturation equation} more relevant. This will be discussed in the next section.

In order to find a closed model for supersaturation, it is necessary to include the equation describing vapour condensation/deposition, $\dd q/\dd t$. At a fixed time, the mixing ratio depends on the concentration, size and density of the particles in the following fashion \cite{rogersyau}:
\begin{equation} \label{eq:mixing ration equation}
    q = \frac{4\pi N}{3\rho _a}\int _0^1\int _{-\infty}^{\infty}\int_0^{\infty}f(r,\rho,c)\rho r^3\dd r\dd \rho \dd,
\end{equation}
where $f$ is the particle size distribution and the different variables are understood for either liquid droplets or ice particles. It is worth noting that the first integral ranging from zero to unity refers to the possible distribution of capacitances, which reflects the efficiency with which droplets/ice particles collect water vapour \cite{rogersyau}. While the capacitance of droplets is approximately $1$--- since all of them are almost spheres---, ice particles possess a wider range of possible shapes that are displayed under certain temperature and humidity conditions, affecting the collection of water vapour. Taking the time derivative of Eq.~\eqref{eq:mixing ration equation}:
\begin{equation}\label{derivative ice mixing ratio}
    \frac{\dd q}{\dd t} = \frac{4\pi N}{\rho _a}\int _0^1\int _{-\infty}^{\infty}\int_0^{\infty}f(r,\rho,c)\rho r^2\frac{\dd r}{\dd t}\dd r\dd \rho \dd c.
\end{equation}
We now employ the diffusional growth law for droplets/particles, which states that water vapour condensates/deposits proportionally to the available supersaturation and inversely proportionally to the radius of the particle \cite{rogersyau}:
\begin{equation}\label{diffusional growth}
    \frac{\dd r}{\dd t}=\frac{cA_{\circ}\SSS}{r}.
\end{equation}
Since we are considering a monodisperse particle size and shape, the previous equation does not only model the diffusional growth of a single particle but of the whole population. Inserting Eq.~\eqref{diffusional growth} into Eq.~\eqref{derivative ice mixing ratio} we find:
\begin{equation}\label{ice mixing ratio derivative averaged}
    \frac{\dd q}{\dd t} = \SSS \frac{4\pi c\rho A_{\circ} N r}{\rho_a}.
\end{equation}
Finally, to obtain the full evolution equation for $\SSS$, the expression for water vapour absorption in Eq.~\eqref{ice mixing ratio derivative averaged} is substituted into Eq.~\eqref{eq:supersaturation equation}. The updraft term is not yet specified, although we anticipate that an SDE for updrafts will be coupled to the supersaturation equation in Section \ref{sect: theory}.

It is clear, then, that a closed equation for the evolution of supersaturation with respect to ice can be obtained by simply integrating Eq.~\eqref{ice mixing ratio derivative averaged} and plugging into Eq.~\eqref{eq:supersaturation equation} to give:
\begin{equation}\label{eq:int_diff}
    \frac{1}{1+\SSS}\dd \SSS = -b\SSS(t)\sqrt{r(0)^2 + 2cA_{\circ}\int _{0}^t\SSS(s)\dd s}\dd t + au.
\end{equation}
The solution of this equation has been shown to converge to a steady state value, which is understood as an equilibrium supersaturation, whereby the changes in relative humidity are balanced by the absorption/release of water vapour by the droplets or ice particles \cite{Korolev2003}. Numerical solutions of this equation can be achieved, although its analytical treatment entails expansions and approximations \cite{Devenish2016}. Equation \eqref{eq:int_diff} is a nonlinear integro-differential equation, since it possesses a square-root of a memory or integral term which accounts for all past supersaturation configurations. The memory term is, in principle, only valid for all the time that a droplets grow without evaporating, sedimenting or precipitating. If any of these processes occurs, memory is lost and the Eq.~\eqref{eq:int_diff} would have to be reinitialised. To account for this issue, in Section~\ref{sec:diffusional growth} we shall develop a theoretical framework to understand the net effects of the fluctuations in droplet or particle radius on the evolution of supersaturation.

\subsection{Stochastic model for updrafts}\label{sect: theory}

The aim of this section is to construct a stochastic model for updraft fluctuations. For this, a cloudy parcel is supposed to be embedded in a turbulent environment with a prescribed surrounding supersaturation $\SSS_E$, which is allowed to be negative in case of dry, subsaturated air. The vertical velocity $u$ in such cloud parcel is assumed to be decomposed into its mean $\bar{u}$ and fluctuating part $u'$ so that $u = \bar{u} + u'$. Given the small spatial scales considered herein, turbulence is taken to be isotropic so that the variance of $u'$, $\sigma_u$, is related to the turbulent kinetic energy $\TKE$ of the background flow in the following way:
\begin{equation}
	\TKE = \frac{1}{2}\left( \sigma^2_x + \sigma^2_y + \sigma^2_u \right)  = \frac{3}{2}\sigma^2_u.
\end{equation}
The updraft variance, $\sigma_u^2$, together with the eddy dissipation rate $\varepsilon$ provide an estimate of the exponential rate, $1/\tau_d$, at which the fluctuations $u'$ decorrelate in time. Such number is called the Lagrangian decorrelation timescale \cite{rodean1996}. Roughly speaking $\tau_d$ indicates the amount of time needed for a turbulent flow to forget its original configuration:
\begin{align}
    \tau_d=\frac{2\sigma^2_u}{\epsilon C_0}. \label{eq: rodean formula 2}
\end{align}
Hence, if we assume that vertical motion is homogeneous, random, with stationary mean $\meanu$, variance $\sigma^2_w$ and decorrelation time $\tau_d$, the simplest model for the vertical velocity is the following red noise equation \cite[\S 3.5]{rodean1996}:
\begin{equation}\label{eq:ou_updraft}
    \dd u(t) = -\frac{1}{\tau_d}\left( u(t) - \meanu \right)\dd t +\sqrt{\frac{2}{\tau_d}}\sigma _u \dd W_t,
\end{equation}
where $W_t$ denotes a standard Wiener process, which accounts for the acceleration increments over $\dd t$ time-units, that come from random pressure fluctuations which decorrelate instantly on time. Equation \eqref{eq:ou_updraft} has the structure of an Ornstein-Uhlenbeck (OU) process, for which there exists a vast collection of analytical results \cite{ornstein_1930,pavliotisbook2014}. In particular, this one-dimensional Gaussian process satisfies the following mean, variance and correlation properties:
\begin{subequations}
\begin{align}
    \mathbb{E}\left[u(t)\right] &= e^{-t/\tau_d}u(0) + (1-e^{-t/\tau_d})\bar{u} \\ \mathrm{Var}\left( u(t) \right) &= \sigma_u^2\left( 1 - e^{-2t/\tau_d}\right)  \\ \mathbb{E}\left[u(t)u(0)\right] &= \sigma_u^2e^{-t/\tau_d}. \label{eq:decorr ou}
\end{align}
\end{subequations} 
where $u(0)$ is the initial value of the vertical velocity. Thus, as $t$ tends to infinity, the solution of Eq.~\eqref{eq:ou_updraft} will distribute according to a Gaussian function with mean $\meanu$ and variance $\sigma^2_u$.

As a result of considering a cloud in contact with the exterior, turbulent motion will also be in charge of mixing mass at the cloud edges; see \cite{eytan2022} for a recent discussion of cloud edges and transition zones in cumulus clouds in terms of adiabaticity of their components. This process is assumed to be acting at a constant rate and will drive the supersaturation at the cloud towards an equilibrium value $\SSS_E$. In other words, turbulent mixing rates determine the characteristic time $\tau_{mix}$ to homogenise a cloud volume with is surrounding reservoir \cite{khain_pinsky_2018}:
\begin{equation}
    \tau_{mix}=\left(\frac{L^2}{\epsilon}\right)^{1/3},
\end{equation}
where $L$ is the characteristic length of the turbulent zone. Therefore, a more general form for the supersaturation equation is studied so that mixing at the cloud edges is also taken into account:
\begin{equation}\label{eq:coupled_equations}
    \frac{1}{1+\SSS}\frac{\dd \SSS}{\dd t} = au - b\SSS r - \frac{1}{\tau_{mix}}\left(\SSS - \SSS_E \right).
\end{equation}
Finally, it is enough to couple this equation to Eq.~\eqref{eq:ou_updraft} and the diffusional growth equation \eqref{diffusional growth} to obtain a closed set of equations for the time evolution of supersaturation.

\section{Quasi-steady model statistics}\label{sec:3}

The diffusional growth rate is inversely proportional to the radius of the droplet or particle in question. Therefore, small droplets or particles grow faster compared to larger ones. In this sense, if Eq.~\eqref{diffusional growth} is initialised with a large radius $r^2(0)$, it is expected that it will remain almost constant, at least for the interval where the following inequality is satisfied; see also \cite{Korolev2003,Devenish2016}:
\begin{equation}
    r^2(0) \gg \left| 2cA_{\circ}\int_0^{t}\SSS (s)\dd s \right|.
\end{equation}
Hence, assuming that changes in the size of the cloud droplets can be neglected, we can take $r$ in Eq.~\eqref{derivative ice mixing ratio} to be constant, equal to $\bar{r}$. This is called the quasi-steady approximation \cite{Korolev2003}. The resulting equation loses the time-dependence of the variable $r$ and reads as:
\begin{subequations}\label{eq:coupled_qs_equations}
\begin{align} \label{eq:qs_model}
   \frac{\dd \SSS}{\dd t} &= -(B+C)\left( \SSS - \frac{C\SSS_E}{B+C} \right)(1+\SSS) + a(1+\SSS)u  \\ \dd u &= -\frac{1}{\tau_d}\left( u - \meanu \right)\dd t +\sqrt{\frac{2}{\tau_d}}\sigma _u  \dd W_t,
\end{align}
\end{subequations}
where two constants have been introduced:
\begin{subequations}
\begin{align}
B&=bB_0N\bar{r}; \\
C&=\left(\frac{\epsilon}{L^2}\right)^{1/3}.
\end{align}
\end{subequations}
Note that the radius $r$ is no longer a variable and that is absorbed into the constant $B$. The target of this section is to derive analytically the stationary statistics of model \eqref{eq:coupled_qs_equations} in a variety of approximations that are detailed in the subsections bellow.

\subsection{Fast decorrelation timescale: $\tau_d \ll 1$}\label{sec:fast_decorr}
There are two different timescales involved in Eq.~\eqref{eq:coupled_qs_equations}. One is set by the constant $B+C$, as the characteristic time for the absorption of water vapour, and the other is the Lagrangian decorrelation timescale $\tau_d$. When the Lagrangian decorrelation timescale is small, the turbulent flow takes less time to forget its initial configuration compared to the typical time of approach to the equilibrium of supersaturation. In the limit of $\tau_d\rightarrow 0$, this argument suggests that $u(t)$ will become a stochastic delta-correlated process, i.e., white noise. Indeed, by referring to the theory of homogenisation \cite[Chapter 11, Result 11.1]{pavliotisbook2014}, we are able to, mathematically rigorously, reduce the two-dimensional system describing $\SSS$ and $u$ to:
\begin{equation}\label{supersaturation equation with wiener process}
\dd\SSS = \left[ -(B+C)\left( \SSS - \frac{C\SSS_E}{B+C} \right) \right] (1+\SSS)\dd t+ A\bar{u}(1+\SSS) + A(1+\SSS)  \dd W_t,
\end{equation}
where a new constant $A$ has been introduced and defined as:
\begin{equation}
    A=a\sigma_u\sqrt{2\tau_d},
\end{equation}
which is the normalisation constant necessary to take this diffusion limit; see \cite[\S 6.3]{rodean1996}. Because we have multiplicative noise, we have to specify the stochastic calculus formalism being employed. For simplicity, Eq.~\eqref{supersaturation equation with wiener process} shall be studied under the It\^o formalism, although the Stratonovich viewpoint can be taken instead \cite{gardiner2009}. For completeness, we note that in order to convert the Stratonovich version of Eq.~\eqref{supersaturation equation with wiener process} into It\^o, we apply the It\^o-to-Stratonovich correction $h(\SSS)$ which reads as \cite{gardiner2009,pavliotisbook2014}:
\begin{equation}
h(\SSS) = \frac{A^2}{2} \partial_{\SSS}(1+\SSS)^2 - \frac{A^2}{2}(1+\SSS)\partial_{\SSS}(1+\SSS) = \frac{A^2}{2}(1+\SSS).
\end{equation}
The resulting It\^o stochastic differential equation is:
\begin{equation} \label{eq:red_qs}
\dd\SSS = \left[ -(B+C)\left( \SSS - \frac{2C\SSS_E -A^2}{2(B+C)} \right)  \right] (1+\SSS)\dd t+ A\meanu(1+\SSS)\dd t + A(1+\SSS) \dd W_t.
\end{equation}
Notice that the equilibrium supersaturation $\SSS_E$ is now modified by the term $-A^2$, resulting from the mere presence of multiplicative noise. The best choice of stochastic formalism is not discussed here, although this equation reveals that the full form of the Squire's equation encodes nonlinear interactions between supersaturation and turbulent fluctuations which yield nontrivial corrections to the stochastic formulation of supersaturation evolution.

Because the system is stochastic, the solutions of Eq.~\eqref{supersaturation equation with wiener process} give different results for each noise realisation. To solve this problem, averages over all possible realisations are taken so the system is described in terms of probability distributions. In this section, we shall assume that in a developing cloud the supersaturation budget quickly approaches stationarity, so that the probability of encountering certain value of supersaturation is provided the Fokker-Planck equation \cite{risken} associated with Eq.~\eqref{supersaturation equation with wiener process}:
\begin{equation}\label{eq: fpe supersaturation full}
\partial _{t}f(\SSS,t)=\partial_{\SSS}\left[ (B+C)\left(\SSS-\frac{C\SSS_E+A\bar{u}}{B+C}\right)(1+\SSS)f(\SSS,t) +  \frac{A^2}{2}(1+\SSS)^2\partial_{\SSS}f(\SSS,t) \right],
\end{equation}
where $f(\SSS, t)$ indicates--- when normalised--- the probability of encountering a supersaturation of $\SSS$ at time $t$. Roughtly speaking, this equation says that a density function is advected and diffused by the linear and stochastic components of Eq.~\eqref{supersaturation equation with wiener process}, respectively. 

Unlike the analytical calculations of \cite{Field2014}, the multiplicative noise and nonlinearities involved in Eq.~\eqref{supersaturation equation with wiener process} suggest that the resulting probability densities $f$ cannot be Gaussian. To find it out, we consider the stationary version Eq.~\eqref{eq: fpe supersaturation full}, which, after successive integrations detailed in Appendix~\ref{app:stationary distribution}, the normalised time-independent density is:
\begin{equation}\label{eq: gamma pdf}
    \boxed{f_{1}(\SSS)=\frac{\alpha^{\alpha + \beta + 1}}{e^{\alpha}\Gamma(\alpha + \beta + 1)}e^{-\alpha \SSS}(1+\SSS)^{\alpha + \beta}},
\end{equation}
where we have introduced the nondimensional parameters $\alpha$, $\beta$ and $\SSS^{\ast}$ defined as:
\begin{subequations}
\begin{align}
    \alpha &= \frac{2(B+C)}{A^2} \label{formula variance alpha};\\
    \beta  &= \alpha \SSS^{\ast} \label{eq:beta}; \\
    \SSS ^{\ast} &= \frac{C\SSS_E + A\bar{u}}{B+C}.
\end{align}
\end{subequations}
The yielding moments can be computed accordingly in terms of successive Gamma functions. The general formula for the uncentred moments of this distribution is:
\begin{equation}
    \mathbb{E}\left[  \SSS ^n \right] = \sum _{k=0}^n (-1)^{k+1}\begin{pmatrix}n \\ k\end{pmatrix}\alpha^{-k}\prod _{\ell=1}^{k}(\alpha + \beta + \ell).
\end{equation}
In particular, the formula for the variance is:
\begin{equation}
    \mathrm{Var}\left(\SSS \right) =\mathbb{E}\left[ \left( \SSS - \mathbb{E}\left[ \SSS \right]\right)^2 \right] =  \frac{\alpha + \beta + 1}{\alpha^2}.
\end{equation}
The parameter $\alpha$ is non-dimensional and measures the relative strength of the updraft fluctuations through $A$ and the combination of the absorption and mixing timescales $B+C$. Hence, it follows that a large value of $\alpha \gg 1$ yields a smaller variance. Moreover, in the limit of large $\alpha$, the function $f_1$ can be recast into a Gaussian by means of Laplace's method; see, e.g., \cite{butler_2007}.

As $t$ tends to infinity, the expected value of $\SSS$ will converge to an equilibrium value at a characteristic rate given by the phase relaxation $\tau_p$ \cite{Korolev2003}. If the dynamics where deterministic, such rate would be given by:
\begin{equation}\label{eq:phase_relaxation_nonlinear}
\tau_p = \frac{1}{A\bar{u} + B + C},
\end{equation}
which is obtained by examining the exponent of the solution of the ordinary differential equation \eqref{eq:red_qs}; see details in \cite{Korolev2003}. In the stochastic context, such relaxation rate is obtained by taking the expectation over all noise realisations. However, it would be necessary to solve an open system involving higher order moments of supersaturation \cite{sardina_2015,siewert_bec_krstulovic_2017}. Hence, it is not clear whether Eq.~\eqref{eq:phase_relaxation_nonlinear} is still the phase relaxation for noisy updraft fluctuations. Nevertheless, the solution of the deterministic version of Eq.~\eqref{eq:red_qs} is solved and compared against an ensemble mean of Eq.~\eqref{eq:red_qs}, both cases subject to a mean updraft of $\bar{u}=0.2 \mathrm{m/s}$. Such ensemble mean is calculated by taking $10^4$ noise realisations and averaging them at every time step. The results are plotted in Figure~\ref{fig:phase_relaxation}, where we also include the relaxation curve for the linear approximation that will be explained in the next section. The initial condition of supersaturation is taken to be $0.5$ just for demonstration purposes. It is observed that the deterministic relaxation (orange curve) provides a good estimate for the stochastic one (blue curve), although a slight divergence is seen as time increases.

\subsubsection{Linear approximation}\label{sect: approximate squires equation}

\begin{figure}[h]
    \centering
    \includegraphics[width=0.6\linewidth]{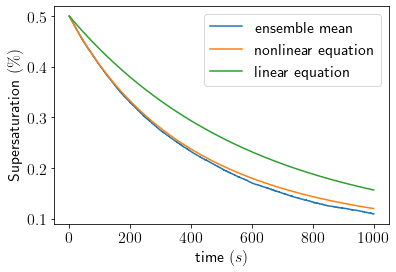}
    \caption{\label{fig:phase_relaxation}\textbf{Phase-relaxation to equilibrium.} The blue curve represents the mean of an ensemble of $10^4$ noise realisations of Eq.~\eqref{eq:red_qs}, with a prescribed mean updraft of $\bar{u}=0.2$ $\mathrm{m/s}$. The orange curve is the solution of Eq.~\eqref{eq:red_qs}, where noise is vanishing. The green curve is the exponential relaxation time obtained from the linear equation \eqref{eq: approx squires equation}. The ambient conditions in this numerical experiment are: $\SSS(0) = 0.5$, $T = -10^{\circ}$, $p = 50500\mathrm{Pa}$, $\bar{r} = 5 \mu\mathrm{m}$ and $N_i = 100\mathrm{L}^{-1}$.}
\end{figure}

For stratocumulus and cumulus clouds, supersaturation does not go beyond $0.5\%$, so that the approximation $1+\SSS \approx 1$ is widely taken in the literature as a first order approximation of the Squire's equation \cite{khain_pinsky_2018}. Under this approximation and some algebraic manipulations, Eq.~\eqref{supersaturation equation with wiener process} becomes an OU process:
\begin{equation}\label{eq: approx squires equation}
   \dd \SSS = -(B+C)\left(\SSS - \SSS^{\ast}  \right)\dd t + A\dd W_t.
\end{equation}
Such linear stochastic differential equation describes a process which has an invariant density equal to a Gaussian distribution. In particular, mean and variance are given by:
\begin{subequations}
\begin{align}
	\mathbb{E}\left[ \SSS (t) \right] &= e^{-(B+C)t}\SSS(0) + \SSS^{\ast}\left(1 - e^{-(B+C)t} \right) \label{eq:mean_linear} \\
	\mathrm{Var}\left(\SSS (t) \right) &= \frac{1}{\alpha}\left( 1 - e^{-2(B+C)t} \right)= \frac{A^2}{2(B+C)}\left( 1 - e^{-2(B+C)t} \right)
\end{align}	
\end{subequations}
As $t$ tends to infinity, the supersaturation value subject to a constant mean updraft tend to an equilibrium value $\SSS^{\ast}$ exponentially fast, with rate given by $-(B+C)$. Hence, the steady-state statistics are provided by a Gaussian distribution:
\begin{equation}\boxed{\label{eq: gaussian pdf}
    f_2(\SSS)=\frac{\alpha}{ \sqrt{2\pi}}e^{-\frac{\alpha(\SSS - \SSS^{\ast})^2}{2}}}.
\end{equation}
Unlike the nonlinear supersaturation equation investigated in Section~\ref{sec:fast_decorr}, the phase relaxation for the present linear version yields the same relaxation time as its deterministic analogue. This is given by:
\begin{equation}\label{eq: relaxation time gaussian}
    \tau _p=\frac{1}{B+C}.
\end{equation}
Note that the phase relaxation in Eq.~\eqref{eq:phase_relaxation_nonlinear} is different to Eq.~\eqref{eq: relaxation time gaussian}, in that the mean updraft $\bar{u}$ is not present in the latter. Consequently, it is only in the limit of $\bar{u}\approx 0$ when both equations converge to equilibrium at the same rate. For illustration, Eq.~\eqref{eq:mean_linear} is plotted in Figure~\ref{fig:phase_relaxation} in green colour, and demonstrates that the nonlinearity accelerates the convergence to equilibrium.

\subsection{Slow decorrelation timescale: $\tau_d\gg 1$}\label{sec:long decorrelation time-scale}

\begin{figure}[h]
    \centering
    \includegraphics[width=0.6\linewidth]{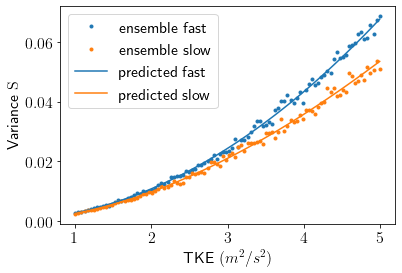}
    \caption{\label{fig:1}\textbf{Supersaturation variance as a function of TKE}. The blue colours correspond to the variances for the 2D linear system \eqref{eq:slow_equation} with red-noise updrafts, whereas the black colours correspond to the homogenised 1D equation \eqref{eq: approx squires equation}, where updrafts become white in time. The dots are obtained by calculating the variance of $10^6$-second time series with a time-step of $10^{-2}$ seconds, for each value of $\TKE$. The solid lines are the predicted variance using the formulas \eqref{formula variance alpha} and \eqref{eq:formula_variance_slow}, for the blue and orange curves, respectively. We highlight that when turbulence is less energetic, both estimates become more similar. The ambient conditions in this numerical experiment are: $\SSS_E = 0$, $\bar{u}=0$, $T = -10^{\circ}$, $p = 50500\mathrm{Pa}$, $\bar{r} = 5 \mu \mathrm{m}$ and $N_i = 100\mathrm{L}^{-1}$.}
\end{figure}

When there is no timescale separation between the phase-relaxation and the updraft fluctuations, the limit of $\tau_d$ going to zero cannot be taken and therefore the white noise limit of the the previous section is not valid. Because of this, the resulting nonlinear equation posses invariant statistics that are intractable analytically. Hence, we start from Eq.~\eqref{eq:coupled_equations} and assume, as in Section~\ref{sect: approximate squires equation}, that $1 + \SSS \approx 1$. We obtain the following  two-dimensional stochastic linear equation:
\begin{subequations}\label{eq:slow_equation}
	\begin{align}
	\dd \SSS &= -(B+C)\left(\SSS - \frac{C\SSS_{E}}{B+C}\right)\dd t + au\dd t, \\ \dd u &= -\frac{1}{\tau_d}\left(u - \bar{u}\right)\dd t + \sqrt{\frac{2}{\tau_d}}\sigma_u\dd W_t.
	\end{align}
\end{subequations}
Which can be compactly recast into matrix form:
\begin{equation}
	\dd \begin{bmatrix} \SSS \\ u \end{bmatrix} = \mathbf{B}\left(\begin{bmatrix} \SSS \\ u \end{bmatrix} - \mathbf{m} \right)\dd t + \mathbf{A}\dd \mathbf{W}_t,
\end{equation}
where $\mathbf{W}_t$ is a two-dimensional independent Wiener process and where the matrices $\mathbf{B}$ and $\mathbf{A}$ and the vector $\mathbf{m}$ are defined as:
\begin{subequations}
	\begin{equation}
		\mathbf{B}=\begin{bmatrix} b_{11} & b_{12}\\ 0& b_{22}\end{bmatrix}=\begin{bmatrix} -(B+C) & a\\ 0& -1/\tau_d\end{bmatrix},
	\end{equation}
	\begin{equation}
		\mathbf{A}=\begin{bmatrix} 0 & 0 \\ 0& a_{22}\end{bmatrix}=\begin{bmatrix} 0 & 0 \\ 0& (2/\tau_d)^{1/2}\sigma_u\end{bmatrix},
	\end{equation}
		\begin{equation}
	\mathbf{m}=\mathbf{B}^{-1}\begin{bmatrix}
C\SSS_E \\ \frac{\bar{u}}{\tau_d}
	\end{bmatrix} = \begin{bmatrix} \frac{C\SSS_E + a\bar{u}}{B+C} \\ \bar{u} \end{bmatrix}.
	\end{equation}
\end{subequations}
Notice that the present equation is degenerate--- noise affects directly to only one variable---, although it will posses an invariant distribution wich possesses a smooth density function. However, the two-dimensional covariance matrix cannot be obtained straightforwardly out of the noise law, but by computing the following matrix integral \cite[Proposition 3.5]{pavliotisbook2014}:
\begin{equation}
	\Sigma = \begin{bmatrix} \sigma_{11} & \sigma_{12} \\ \sigma_{12} & \sigma_{22} \end{bmatrix}= \int _0^{\infty} e^{\mathbf{B}s} \mathbf{A}\mathbf{A}^{\top}e^{\mathbf{B}^{\top}s} \dd s = \begin{bmatrix}
\frac{a^2\sigma_u^{2}}{2(B+C)(B+C+1/\tau_d)} & \frac{-a\sigma_u^{2}}{2(B+C+1/\tau_d)} \\ \frac{-a\sigma_u^{2}}{2(B+C+1/\tau_d)}  & \sigma_u^2
	\end{bmatrix} .
\end{equation}
The resulting process possesses, as in the one-dimensional case, a Gaussian stationary distribution where now:
\begin{equation}
	\boxed{f_{3}(\SSS,u)=(2\pi)^{-1}\left(\mathrm{det}\Sigma\right)^{-1/2}e^{-\frac{1}{2}\left([\SSS,u] - \mathbf{m}^{\top}\right)\Sigma^{-1}\left([\SSS,u]^{\top} - \mathbf{m}\right)}}.
\end{equation}
To derive the marginal distribution for $\SSS$ we apply the affine transformation $P=[1,0]$ to the random variable $[\SSS,u]^{\top}$ so that $P[\SSS,u]^{\top} = \SSS$. Hence, 
\begin{equation}\label{eq:formula_variance_slow}
P\begin{bmatrix}
\SSS \\ u
\end{bmatrix} = \SSS \sim \mathcal{N}\left( P\mathbf{m},P\Sigma P^{\top}\right) = \mathcal{N}\left( \frac{C\SSS_E + a\bar{u}}{B+C},\frac{a^2\sigma_u^{2}}{2(B+C)(B+C+1/\tau_d)}\right)
\end{equation}
The timescale separation argument to reduce the equation for the evolution of supersaturation is only valid for small values of $\TKE$. This hypothesis is tested and shown in Figure~\ref{fig:1}, where the variance of supersaturation in a cold cloud is computed using the formulas here presented and its numerical estimation using long time series of $10^{6}$ seconds. The equations are integrated using a simple Euler-Maruyama method with a time-step of $10^{-2}$ seconds \cite{gardiner2009}.

\section{Diffusional growth and size fluctuations}\label{sec:diffusional growth}

In the Introduction we presented the full and closed equation for the evolution of ice supersaturation accounting for the condensational growth of droplets/ice crystals coupled to turbulent updraft fluctuations; see Eq.~\eqref{eq:int_diff}. In Section~\ref{sec:3}, we investigated the properties of the stochastic Squires equation, describing the evolution of supersaturation in the quasi-steady case, where condensational growth is neglected. Here, we wish to explore the effects of turbulent random forcings on the diffusional growth and the latter's on supersaturation evolution. Two default approximations are taken in this section. First, we will assume that updraft fluctuations decorrelate instantly so that the white-noise model is valid. Secondly, the small-supersaturation approximation is taken: $1+\SSS \approx 1$. Then, by applying the chain rule to Eq.~\eqref{diffusional growth}, we can write the supersaturation equation coupled to diffusional growth as:

\begin{subequations}\label{eq:fully_coupled}
\begin{align}\label{eq:diff_r2}
\dd \SSS &= -C(\SSS - \SSS^{\ast}_d)\dd t  -B_dr\SSS \dd t + A\dd W_t \\ 
\dd r^2 &= F\SSS \dd t,\text{ }r^2>0 \label{eq:chain_r2} \\
\dd r^2 &= 0,\text{ }r^2=0, \text{ and } \SSS(t)<0. \label{eq: diff growth perturbations 2 3}
\end{align}
\end{subequations}
Where some constants are introduced:
\begin{subequations}
\begin{align}
\SSS^{\ast}_d &= \frac{C\SSS_E + A\bar{u}}{C}; \\
B_d &= bB_0N; \\
F &= 2cA_{\circ}.
\end{align}
\end{subequations}
We note that the coupling of the equations \eqref{eq:diff_r2} and \eqref{eq:chain_r2} is nonlinear since it involves the square-root of the variable $r^2$ and makes the analytics intractable. However, we refer at this stage to Appendix~\ref{app:square_root} for the analysis of the square-root stochastic process.

When the nonlinear term $r\SSS $ is small and close to zero, the mean and variance of $r^{2}$ evolve according to:
\begin{subequations}
\begin{align}
\mathbb{E}\left[ r^2(t)\right]&=  F\int ^t\mathbb{E}\left[\SSS(s) \right] \dd s + r^2(0) \approx \frac{Fe^{-Ct}}{C}\left(\SSS^{\ast}_d - \SSS(0)\right) +F\SSS^{\ast}_dt + r^2(0) \\ \label{eq: average sequared displacement}
\mathbb{E}\left[ \left( r^2(t)\right)^2\right] &= F^2\mathbb{E}\left[ \int_0^t\int_0^t\SSS (s)\SSS(u)\dd s \dd u \right] = 2F^2 \int _0^t\int _0^u \frac{A^2}{2C}\left( e^{-C(u-s)} - e^{-C(u+s)} \right) \dd s \dd u \\ \label{eq: analytical spread diffusional growth}&= \frac{F^2A^2}{C^2}t - \frac{F^2A^2}{2C^3}+\frac{F^2A^2}{2C^3}\left( 4e^{-Ct}-e^{-2Ct} \right).
\end{align}
\end{subequations}
These set of formulas are identical to those obtained in the 30s in the study of Brownian motion--- see \cite{ornstein_1930}---, and imply that the evolution of the squared radius follows Brownian paths where, in particular, the mean-squared displacement scales linearly for large times, if negative values of squared radii were allowed. Indeed, if $t\gg1$ and the nonlinear coupling is small:
\begin{equation}\label{eq:variance approximation}
\mathbb{E} \left[\left( r^2(t)\right)^2\right] \approx \frac{F^2A^2}{C^2}t.
\end{equation}
The boundary condition $r^2 = 0$ is strictly necessary since it is possible that trajectories of $r(t)$ in Eq.~\eqref{eq:chain_r2} vanish. When that happens, it means that the particles or droplets in question have evaporated and that the formula for the variance in Eq.~\eqref{eq: analytical spread diffusional growth} has to be reinitialised once the droplets and particles have reactivated.

While under this framework there is no stationary distribution with finite variance for particle radius, it was shown in \cite{siewert_bec_krstulovic_2017} that if the ambient supersaturation $\SSS_E$ is negative, i.e., subsaturated, the probability distribution of $r^2$ will possess the structure of an exponential function with an Dirac-peak located at $r^2=0$ which arises from the boundary condition in Eq.~\eqref{eq: diff growth perturbations 2 3}.

\subsection{Fluctuations in droplet radius}
In the previous section we clarified that if supersaturation is let to be driven by random turbulent updrafts, the mean-square radius grows linearly in time and, therefore, unbounded Brownian excursions can be expected when solving the condensational growth equation. When the cloud in question is in contact with a subsaturated environment or the system allows for droplet evaporation or sedimentation, it is expected that the trajectories in the $\SSS$-$r^2$-plane of Eq.~\eqref{eq:fully_coupled} will display cycles, where $r^2$ grows but then vanishes and sticks at the boundary of $r^2=0$ for an open interval of time; see \cite[Figure 4]{siewert_bec_krstulovic_2017}. In this section, we aim at calculating the net effects of the mentioned cycles, by modelling them as an extra random forcing at the microphysical term. For this purpose we introduce a new constant $\sigma_r$ which indicates the standard deviation of droplet radius fluctuations. For small particles where condensation is a dominating growth factor, $\sigma_r$ must be proportional to $F$ and the standard deviation of $\SSS$. To simplify the expression, we shall present the results for $\SSS_E = 0$ and $\bar{u}=0$.

The starting point is the $1+\SSS \approx 1$ approximation that together with fluctuations in droplet size lead to a new model for supersaturation:
\begin{equation}\label{eq:sto_droplet}
    \dd \SSS (t) = -C\SSS(t)\dd t - B_d \SSS(t)\left(\bar{r}\dd t + \sigma_r\dd W^{(1)}_t\right) + A\dd W^{(2)}_t,
\end{equation}
where $W^{(1)}$ and $W^{(2)}$ are two independent Wiener processes. Also, the parameter $B$ has been modified to $B = bB_0N$. In this case, solving the stationary Fokker-Planck equation for $\SSS^{\ast} = 0$ yields the following non-normalised stationary distribution:
\begin{equation}\label{eq:dist4}
    \boxed{f_4(\SSS) = \left(A^2 + \sigma_r^2B_d^2\SSS^2 \right)^{-1-\frac{C+B_d\bar{r}}{B_d^2\sigma_r^2}}}.
\end{equation}
This formula constitutes a power-law, for which higher momenta might not be well defined. However, by examining the exponent, it follows that the distribution $f_4$ has $M$ momenta, if $M/2<1+(C+B_d\bar{r})/B_d^2\sigma_r^2$. The function $f_4$ describes the steady-state statistics when $W^{(1)}_t \neq W^{(2)}_t$, but since fluctuations in droplets size typically originate in fluctuations of supersaturation, it is important to consider the the case with $W^{(1)}_t = W^{(2)}_t$, so that noise sources are correlated. In such scenario, the stationary distribution is:
\begin{equation}\label{eq:dist5}
    \boxed{f_5(\SSS) = e^{-\frac{2(C+B_d\bar{r})A}{B_d^2\sigma_r^2}(A - B_d\sigma_r\SSS)^{-1}}\left(A - B_d\sigma_r\SSS\right)^{-2-\frac{2(C+B_d\bar{r})}{B_d^2\sigma_r^2}}}.
\end{equation}
The probability distributions of Eq.~\eqref{eq:dist4} and \eqref{eq:dist5} differ from a Gaussian distribution since the noise appears in a multiplicative way. It is clear, though, that the stochastic process Eq.~\eqref{eq:sto_droplet} will converge to linear evolution of supersaturation as $\sigma _r$ tends to zero, which yields the Gaussian distribution of Eq.~\eqref{eq: gaussian pdf}. It is possible, on the other hand, to show the pointwise convergence of Eq.~\eqref{eq:dist4} and \eqref{eq:dist5} to Eq.~\eqref{eq: gaussian pdf}, by means of Laplace's method; see, e.g., \cite{butler_2007}.

\section{Parametrisation of mixed-phase clouds}\label{sec:parametrisation}

The relevance of the analytical calculations done in the previous sections is discussed here in the context of mixed-phase clouds. The coexistence of liquid water and ice is thermodynamically unstable so that, in freezing temperatures, ice will inevitably grow at the expense of liquid water \cite{bergeron_1935}. Observations, on the other hand, indicate that mixed-phase conditions are not just a transient microphysical state, but that supercooled liquid water can be maintained at cloud top temperatures down to $-40^{\circ}$. The activation of supercooled liquid water in a cloud parcel is in general achieved if the following two criteria are met: (i) the vertical velocity must exceed a threshold value and (ii) the cloud parcel must be lifted to a threshold altitude. Under this context, mixed-phased conditions can be kept for a long term 
\cite{Korolev2008}. When considering an icy cloud, Eq.~\eqref{eq:coupled_equations} has to be interpreted with respect to ice supersaturation $\SSi$.

Under this theoretical setting, \cite{Field2014} proposed a method to estimate the variance of ice supersaturation using a Gaussian probability distribution, here revisited in Section~\ref{sect: approximate squires equation}.
Indeed, when supersaturation with respect to ice exceeds the value at liquid water saturation, liquid water is activated. Supersaturation with respect to ice relates to that of water, $\SSw$, in terms of the ratio of their respective equilibrium vapour pressures:
\begin{equation}\label{formula supersaturation ice water}
    \SSi = \eta(T) \SSw + \eta(T) - 1.
\end{equation}
where $\eta(T)=E_w(T)/E_i(T)$. As a consequence, supersaturation with respect to ice at water saturation $\SSiw$ is given by:
\begin{equation}
    \SSiw=\eta(T) - 1.
\end{equation}
 Hence, the fraction of cloud in mixed-phase conditions and liquid water content are determined by the tails of the probability density functions in $\{ f_k \}_{k=1}^5$ from $\SSiw$ to infinity. Concretely, the fraction of cloud parcel that is taken to be in mixed-phase is the total time $\SSi$ spends above $\SSiw$ or, in other words--- by invoking ergodicity---, the integral of $f_k$ from $\SSiw$ to infinity: 
 \begin{equation}\label{eq:cloud fraction formula}
     \mathrm{C}_{f_k} = \lim_{t\rightarrow \infty} \frac{1}{t}\int_{0}^{t}\mathbf{1}_{\SSi\geq \SSiw}\left(\SSi (s)\right)\dd s = \int_{\SSiw}^{\infty} f_k(\SSi)\dd \SSi,
 \end{equation}
where $\mathbf{1}_{\SSi \geq \SSiw}$ is the characteristic function for values of $\SSi$ larger than $\SSiw$. As noted in \cite{Field2014}, in the Gaussian case the cloud fraction is given explicitly by:
\begin{equation}\label{eq: cloud fraction gaussian}
    \mathrm{C}_{f_2}=\frac{1}{2}\mathrm{erfc}\left( \sqrt{\frac{\alpha}{2}}\left(\SSiw - \SSS^{\ast}\right)\right).
\end{equation}
Moreover, the parcel's LWC is calculated by assuming that ice supersaturation above that with respect to liquid water is converted into droplets. The LWC is, then, estimated by integrating the adjustment formula of Eq.~\eqref{formula supersaturation ice water} to obtain the condensed water amount:
\begin{equation}\label{pdf lwc gaussian}
\left \langle q \right\rangle _k = \lim_{t\rightarrow \infty} \frac{1}{t}\int_{0}^{t} \mathbf{1}_{\SSi \geq \SSiw}(\SSi(s)) \left( \SSi(s) - \SSiw \right)\rho_a q_{si}\dd \SSi = \int_{\SSiw}^{\infty}\left( \SSi - \SSiw \right)f_k(\SSi)\rho_aq_{si}\dd \SSi
\end{equation}
where $\rho_a q_{si}$ is the factor that converts liquid water content to supersaturation values.

\begin{figure}[H]
	\centering
	\begin{subfigure}[b]{0.49\textwidth}
		\centering
		\includegraphics[width=\textwidth]{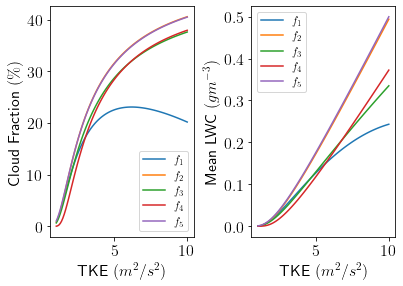}
		\caption{}
	\end{subfigure}
	\hfill
	\begin{subfigure}[b]{0.49\textwidth}
		\centering
		\includegraphics[width=\textwidth]{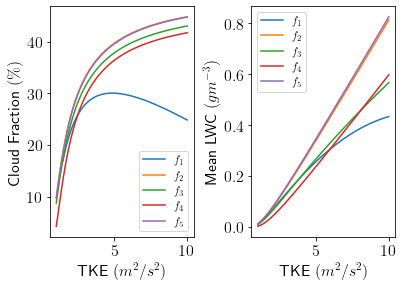}
		\caption{}
	\end{subfigure}
	\caption{\label{fig:2}\textbf{Cloud fraction and mean LWC as a function of TKE.} The cloud fraction and mean LWC are calculated using Eq.~\eqref{eq:cloud fraction formula} and Eq.~\eqref{pdf lwc gaussian}, respectively, against $\TKE$, for each analytical supersaturation distribution $\{ f_k \}_{k=1}^5$ and for two temperature configurations. In (a), a temperature of $-10^{\circ}$ is considered, in (b), $-5^{\circ}$. The rest of the ambient conditions are: $\SSS_E = 0$, $\bar{u}=0$, $\sigma^2_r=10^{-6}\mathrm{m}^2$, $p = 50500\mathrm{Pa}$, $\bar{r} = 5 \mu\mathrm{m}$ and $N_i = 100\mathrm{L}^{-1}$.}
\end{figure}

The five probability distributions obtained in the previous sections are now used to compute the cloud fraction and the mean LWC of a subgrid cloud parcel for a range of free parameters. Such free parameters are the $\TKE$, as a proxy for turbulent forcing, and variance of the droplet radius fluctuations. In Figure~\ref{fig:2}(a) and (b) we show the dependence of the mentioned partial moments on values of $\TKE$ at two temperatures indicated in the captions. Such statistics where computed using a simple quadrature scheme on the interval $[-10,10]$ so that all the considered PDFs integrate to unity with a tolerance of $10^{-10}$. We observe a monotone dependence on $\TKE$ in all the PDFs but for the Gamma distribution, which yielded decreasing cloud fractions for values of $\TKE>6\mathrm{m^2s^{-2}}$ at $-10^{\circ}$ and $\TKE>4\mathrm{m^2s^{-2}}$ at $-5^{\circ}$. Such change in trend is due to the displacement of the mode and tail-thickness in the Gamma distribution as the location factor is altered due to the multiplicative noise. The nonlinear interaction of updraft fluctuations and supersaturation, hence, prevent the increase in mixed-phase conditions for large values of turbulent forcing strength.

\begin{figure}[H]
	\centering
	\begin{subfigure}[b]{0.49\textwidth}
		\centering
		\includegraphics[width=\textwidth]{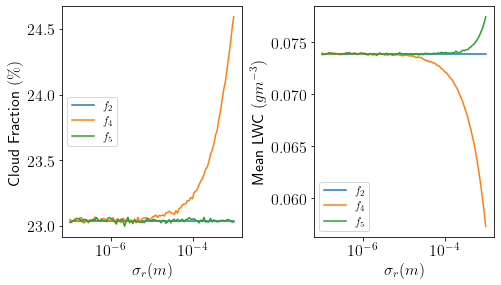}
		\caption{}
	\end{subfigure}
	\hfill
	\begin{subfigure}[b]{0.49\textwidth}
		\centering
		\includegraphics[width=\textwidth]{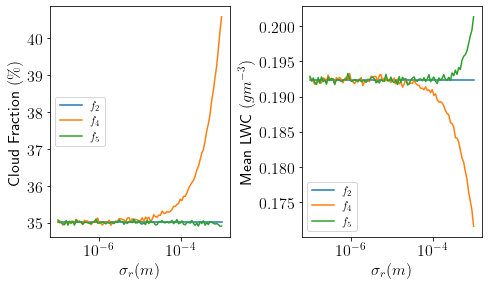}
		\caption{}
	\end{subfigure}
	\caption{\label{fig:3}\textbf{Cloud fraction and mean LWC as a function of $\sigma_r$.} The cloud fraction and mean LWC are calculated using Eq.~\eqref{eq:cloud fraction formula} and Eq.~\eqref{pdf lwc gaussian}, respectively, against $\sigma_r$, for the analytical distributions $f_1,f_4$ and $f_5$ and for two temperature configurations. For small values of $\sigma_r$, the functions $f_4$ and $f_5$ are computationally expensive to evaluate so ergodic averages of $10^{6}$ seconds are taken instead. In (a), a temperature of $-10^{\circ}$ is considered, in (b), $-5^{\circ}$. The rest of the ambient conditions are: $\SSS_E = 0$, $\bar{u}=0$, $\TKE=2.612\mathrm{m^2/s^2}$, $p = 50500\mathrm{Pa}$, $\bar{r} = 5 \mu\mathrm{m}$ and $N_i = 100\mathrm{L}^{-1}$.}
\end{figure}

When droplet size fluctuations are allowed, the first three calculated distributions $f_1,f_2$ and $f_3$ naturally yield the same statistics for cloud fraction and mean LWC. Contrarily, when noisy variations in radius are allowed the distributions of $f_4$ and $f_5$ are likely to display a dependence on $\sigma_r$. This dependence is shown in Figure~\ref{fig:3}, where the cloud fraction and LWC are calculated as a function $\sigma_r$, for $f_2,f_4$ and $f_5$. Because $f_4$ and $f_5$ are expensive to evaluate at small values of $\sigma_r$, ergodic averages are computed instead, i.e., the first equalities in Eq.~\eqref{eq:cloud fraction formula} and \eqref{pdf lwc gaussian}. Such averages are taken over an integration of Eq.~\eqref{eq:sto_droplet} over $10^6$ seconds. While $f_2$ is expectedly constant, $f_4$ and $f_5$ start to be dependent for values larger than $\sigma^2_r\geq 10^{-12}\mathrm{m}^2$, which corresponds to a standard deviation $20\%$ of the mean droplet radius $\bar{r}$. In particular, it is observed that $\mathrm{C}_{f_4}$ grows up to $1\%$ when droplet fluctuations have a variance of $10^{-8}\mathrm{m}$. On the other hand, $\mathrm{C}_{f_5}$ appears to be independent of $\sigma_r$. We recall that this case corresponds to when the fluctuations in $r$ are independent of the noisy updraft. Such independence does not hold in LWC, where, $\langle q \rangle_5$ appears to grow, as a consequence of the fattening of the tails.

\section{Discussion}\label{sec:discussion}

In this paper, the analysis of the stochastic Squires equation has been done, with the aim of providing analytical formulas for the distribution of supersaturation in a cloud parcel. Such equation describes the evolution of supersaturation over time by, essentially, taking into account the sources and sinks of relative humidity due to (a), adiabatic cooling via turbulent updrafts and (b), water vapour condensation onto droplets/particles. The former are here modelled as a red-noise process, in accordance with the spatial and time correlations of isotropic fluctuations in the inertial regime \cite{rodean1996}. Secondly, droplet growth by vapour condensation is studied and noted--- as firstly done in \cite{siewert_bec_krstulovic_2017}--- that the variance of squared droplet radius grows linearly in time and, hence, it does not have bounded steady-state statistics. Here, we propose to include condensational growth effects by adding an explicit stochastic term for droplet size fluctuations which yield analytically tractable probability distributions for supersaturation. In total, a number of five different probability distributions where computed.

In the present paper, the theoretical results of \cite{Field2014}--- here condensed in Section~\ref{sect: approximate squires equation}--- have been generalised to a wider range of contexts. First of all, the Squires equation is considered in Section~\ref{sec:2} in its full nonlinear version and has been shown to possess a Gamma-like stationary distribution--- here denoted as $f_1$--- that deviates from a Gaussian according to the parameter $\alpha$. Such parameter, also found in \cite{Field2014}, is a non-dimensional ratio between the strength of turbulent fluctuations and the phase relaxation coefficient. Thus, as turbulent fluctuations decrease in variance (relative to the microphysical or mixing timescale), $f_1$ becomes better and better approximated by the Gaussian distribution $f_2$. Indeed, that can be seen from the computation of partial moments--- cloud fraction and mean LWC--- in Figure~\ref{fig:2}.

The timescale separation assumption, needed to compute $f_1$ and $f_2$, is lifted in Section~\ref{sec:long decorrelation time-scale}. Indeed, $f_2$ is only valid when times are much greater that the Lagrangian decorrelation timescale so that updraft fluctuations become white in time. The distribution $f_3$ is still Gaussian, but it does not invoke such timescale separation between updraft fluctuations and supersaturation. On Figure~\ref{fig:1}, we show that the variance predicted by $f_2$ and $f_3$ diverge as $\TKE$ increases.

The main assumption needed to compute $f_1,f_2$ and $f_3$ is the quasi-steady approximation, whereby the droplet or ice particle radius is considered constant. It was shown in \cite{siewert_bec_krstulovic_2017} that the long-term variance of droplet squared-radius scales linearly in time, similar to Brownian motion. This is revisited here in Section~\ref{sec:diffusional growth}. We argue that such result is only valid for short times, since large families of droplets are subject to processes like sedimentation, evaporation or mixing with exterior dry air that provoke a \emph{memory loss} in collective droplet growth and, hence, the variance is reinitialised. In this work, instead, we investigate random fluctuations in droplet size in two contexts: (i) fluctuations in updrafts are uncorrelated to those of droplet size and (ii), the source of noise is the same, albeit with different intensities. The net effect of droplet radius fluctuations in summarised in Figure~\ref{fig:3}, where the cloud fraction and mean LWC are computed as a function of $\sigma_r$. It is found that correlated radius and updraft fluctuations yield a more sensitive probability distribution, which deviates severely from the quasi-steady approximation by up to $+5\%$ in cloud fraction when $\sigma_r\geq 10^{-6}\mathrm{m}$. Contrarily, the mean LWC is negatively correlated with fluctuations in $\sigma_r$ in case of $f_4$. Regarding the uncorrelated sources of noise, the cloud fraction appears to be weakly dependant in $\sigma_r$. On the other hand, mean LWC correlates positively is droplet size fluctuation variance.

The present stochastic analysis of supersaturation is argued to be a useful framework to find probabilistic formulas for the parametrisation of subgrid cloud properties. However, a deeper investigation of Eq.~\eqref{eq:int_diff} would be useful in this stochastic framework. One step forward would be to impose a characteristic time for the loss of memory, so that the integro-differential equation can be replaced by a simpler expression, possibly some form of noise with suitable time-decorrelation properties. We anticipate that this would entail technical difficulties due to the square-root nonlinearity--- here minimally tackled in Appendix~\ref{app:square_root}---, so research should be oriented in this direction. In general, we belief that this approach can be extended to more general contexts, possibly, by including more microphysical processes that affect the growth of liquid droplets or ice particles and, hence, the overall regulation of a cloud supersaturation budget.

\section*{Data availability}

The data to produce the figures in this paper is obtained from the numerical integration of the equations in question. The corresponding code is available upon request.

\section*{Acknowledgements}

The authors would like to thank S. Roncoroni, P. Field and B. Devenish for their comments, suggestions and kind reception at the MetOffice UK. MSG is grateful to the Mathematics of Planet Earth Centre for Doctoral Training (MPE CDT) for making this collaboration possible. MSG acknowledges and is grateful for the support of the Institute of Mathematics and its Applications (grant number: SGS21/08). MSG is thankful to I. Koren, M. D. Chekroun, the cloud physics group and the graduate school at the Weizmann Institute of Science for providing a most inspiring environment.

\pagebreak

\begin{appendices}
\numberwithin{equation}{section}

\section{List of some used notations and symbols}
\begin{table}[H]\renewcommand{\arraystretch}{1.5}
    \centering
    \begin{tabular}{lcl}
    \hline
      Symbol & Units &  Description \\
         \hline
        $A$ &  [$\mathrm{s}^{-1/2}$]& $a_i\sigma_u\tau_d^{1/2}$  \\
        $A_i$ & [$\mathrm{m}^{2}\mathrm{s^{-1}}$] & $\left( \frac{\rho_iL_i^2}{kR^vT^2}+\frac{\rho_iR_vT}{E_iD} \right)^{-1}$\\
        $A_w$ & [$\mathrm{m}^{2}\mathrm{s^{-1}}$] & $\left( \frac{\rho_wL_w^2}{kR^vT^2}+\frac{\rho_wR_vT}{E_wD} \right)^{-1}$\\
        $A_{\circ}$ & [$\mathrm{m}^{2}\mathrm{s^{-1}}$] & $A_i$ or $A_w$ depending on the context\\
        $a_w$ & [$\mathrm{m}^{-1}$] & $\frac{g}{R_aT}\left( \frac{L_wR_a}{c_pR_vT} - 1 \right)$\\
        $a_i$ & [$\mathrm{m}^{-1}$] & $\frac{g}{R_aT}\left( \frac{L_iR_a}{c_pR_vT} - 1 \right)$\\  
        $a$ & [$\mathrm{m}^{-1}$] & $a_i$ or $a_w$ depending on the context \\ 
        $B$ & $[\mathrm{s}^{-1}]$ & $b_iB_0N\bar{r}$\\
        $B_d$ & $[\mathrm{r^{-1}~s}^{-1}]$ & $b_iB_0N$\\
        $b_w$ & $[-]$ &$\frac{1}{q_v}+\frac{L_w^2}{c_pR_vT^2}$ \\
        $b_i$ & $[-]$ &$\frac{1}{q_v}+\frac{L_i^2}{c_pR_vT^2}$ \\
        $b_m$ & $[-]$ &$\frac{1}{q_v}+\frac{L_iL_w}{c_pR_vT^2}$ \\
        $b$ & $[-]$ &$b_w$, $b_i$ or $b_m$ depending on the context \\
        $B_0$ & $[\mathrm{m}^{2}\mathrm{s^{-1}}]$ & $\frac{4\pi \rho_iA_ic}{\rho_a}$ \\
        $C_0$ & $[-]$ & Lagrangian structure function constant($=10$)\\
        $C_f$ & $[-]$ & Mixed-phase cloud fraction \\
        $c$ & $[-]$ & Ice-crystal capacitance, shape factor ($=1$) \\
        $c_p$ & $[\mathrm{J}\mathrm{kg}^{-1}\mathrm{K}^{-1}]$ & Specific heat capacity of moist air at constant pressure \\
        $D$ & $[\mathrm{m}^{2}\mathrm{s^{-1}}]$ & Water-vapour diffusion coefficient in air \\
        $e$ & $[\mathrm{Pa}]$ & water vapour pressure \\
        \hline
    \end{tabular}
    \caption{List of symbols.}
    \label{tab: configuration}
\end{table}

\begin{table}[H]\renewcommand{\arraystretch}{1.5}
    \centering
    \begin{tabular}{lcl}
    \hline
      Symbol & Units &  Description \\
         \hline
         $E_i$ & $[\mathrm{Pa}]$ & Saturation vapour pressure over ice \\
        $E_w$ & $[\mathrm{Pa}]$ & Saturation vapour pressure over liquid water \\
        $E$ & $[\mathrm{Pa}]$ & Saturation vapour pressure over liquid or ice depending on the context \\
        $\varepsilon$ & $[\mathrm{m}^2\mathrm{s}^{-3}]$ & Eddy dissipation rate \\
        $F$ & $[\mathrm{m}^2\mathrm{s}^{-1}]$ & $2cA_\circ$ \\
        $g$ & $[\mathrm{m}\mathrm{s^{-2}}]$ & Acceleration due to gravity \\
        $k$ & $[\mathrm{J}\mathrm{m}^{-1}\mathrm{s}^{-1}\mathrm{K}^{-1}]$ & Heat conductivity coefficient in air \\
        $L$ & $[\mathrm{m}]$ & Vertical length of turbulent zone \\
        $L_i$ & $[\mathrm{J}\mathrm{kg}^{-1}]$ & Latent heat for sublimation plus melting \\
                 $L_w$ & $[\mathrm{J}\mathrm{kg}^{-1}]$ & Latent heat for vaporisation \\
        $N_i$ & $[\mathrm{L}^{-1}]$ & Concentration of ice particles \\
        $N_w$ & $[\mathrm{L}^{-1}]$ & Concentration of water droplets \\
        $N$ & $[\mathrm{L}^{-1}]$ & Concentration of water droplets/ice particles depending on the context \\
        $p$ & $[\mathrm{Pa}]$ & Pressure \\
        $\langle q \rangle$ & $[\mathrm{kg~m}^{-3}]$ & Domain mean liquid water content \\
        $q_{si}$ & $[\mathrm{kg~kg}^{-1}]$ & Mass mixing ratio of vapour at ice saturation \\
        $q_v$ & $[\mathrm{kg~kg^{-1}}]$ & Mass mixing ratio of water vapour \\
        $q$ & $[\mathrm{kg~kg^{-1}}]$ & Liquid water or ice mixing ratio depending on the context \\
        $R_v,R_a$ & $[\mathrm{J~kg^{-1}~K^{-1}}]$ & Specific gas constant of water vapour and air, respectively \\

        \hline
    \end{tabular}
    \caption{List of symbols.}
    \label{tab: configuration}
\end{table}

\begin{table}[H]\renewcommand{\arraystretch}{1.5}
    \centering
    \begin{tabular}{lcl}
    \hline
      Symbol & Units &  Description \\
         \hline
        $\SSiw$ & $[-]$ & $E_w/E_i - 1$  \\
        $\SSS_w$ & $[-]$ & Supersaturation with respect to liquid water \\
        $\SSi$ & $[-]$ & Supersaturation with respect to ice \\
        $\SSS$ & $[-]$ & Supersaturation with respect to liquid water or ice depending on the context \\
        $\SSS_E$ & $[-]$ & Environmental supersaturation with respect to liquid water or ice  \\
        $\SSi^{\ast}$ & $[-]$ & $\SSiw + q/(\rho_aq_{si})$ \\
        $T$ & $[\mathrm{K}]$ & Temperature \\
        $\TKE$ & $[\mathrm{m^2/s^2}]$ & Turbulent kinetic energy \\
        $u$ & $[\mathrm{m~s^{-1}}]$ & Vertical velocity \\
        $\bar{u}$ & $[\mathrm{m~s^{-1}}]$ & Mean vertical velocity \\
        $\epsilon$ & $[\mathrm{m^2s^{-3}}]$ & Eddy dissipation rate \\
        $\nu$ & $[\mathrm{m^2~s^{-1}}]$ & Kinematic viscosity \\
        $\rho_i$ & $[\mathrm{kg~m^{-3}}]$ & Effective density of ice \\
        $\rho_a$ & $[\mathrm{kg~m^{-3}}]$ & Density of air \\
        $r$ & $[\mathrm{m}]$ & Droplet or ice particle radius depending on the context \\
        $\bar{r}$ & $[\mathrm{m}]$ & Mean droplet or ice particle radius depending on the context \\
        $\sigma_u$ & $[\mathrm{m~s^{-1}}]$ & Standard deviation of vertical velocity fluctuations \\
        $\sigma_r$ & $[\mathrm{m}]$ & Standard deviation of droplet/particle size fluctuations \\
        $\tau_d$ & $[\mathrm{s}]$ & Lagrangian decorrelation timescale \\
        \hline
    \end{tabular}
    \caption{List of symbols.}
    \label{tab: configuration}
\end{table}

\section{Stationary supersaturation distribution}\label{app:stationary distribution}

The calculation of the stationary distributions of each case study is done by studying the Fokker-Planck representation of the stochastic processes in question \cite{risken}. Such equation describes how probability distributions evolve in time towards its stationary state. For simplicity, in this appendix we will just show how to derive, step by step, the distribution for ice supersaturation Eq.~\eqref{eq: gamma pdf}. In this case, the Fokker-Planck equation associated with Eq.~\eqref{supersaturation equation with wiener process} is given by Eq.~\eqref{eq: fpe supersaturation full}. Then, its stationary version, for $\SSS_E = \bar{u} = 0$, reads as:
\begin{equation}\label{eq: stationary fpe supersaturation full}
0=\partial_{\SSS}\left[ (B+C) \SSS(1+\SSS)f(\SSS) +  \partial_{\SSS}\left(\frac{A^2}{2}(1+\SSS)^2f(\SSS)\right) \right],
\end{equation}
where the time-dependence has been dropped. Integrating from $\ell$ (yet to be determined) to $\SSS$ and assuming that the stationary distribution $f$ and $\partial_{\SSS}f$ vanish at $\ell$:
\begin{align}
    0= (B+C) \SSS(1+\SSS)f(\SSS) + A^2(1+\SSS)f(\SSS) + \frac{A^2}{2}(1+\SSS)^2\partial_{\SSS}f(\SSS).
\end{align}
We now rearrange the equation to make it homogeneous on both sides:
\begin{equation}
\frac{\partial_{\SSS} f(\SSS)}{f(\SSS)}=-\frac{2(B+C)\SSS}{A^2(1+\SSS)} - \frac{2}{1+\SSS}.
\end{equation}
Integrating on both sides:
\begin{equation}
    \mathrm{log}(f(\SSS))=\frac{2(B+C)}{A^2}\mathrm{log}(1+\SSS) - \frac{2(B+C)}{A^2}\SSS - 2\mathrm{log}(1+\SSS).
\end{equation}
Taking exponentials on both sides:
\begin{equation}
f(\SSS) = e^{-\frac{2(B+C)}{A^2}\SSS}(1+\SSS)^{-2+\frac{2(B+C)}{A^2}}.
\end{equation}
This is a non-normalised solution for the stationary Fokker-Planck equation, which has a singularity at $-1$. As $\SSS$ tends to infinity, $f(\SSS)$ goes to zero asymptotically. We now calculate the normalisation constant $C$, for which we employ the parameter $\alpha$ like in Eq.~\eqref{formula variance alpha}:
\begin{align}
    C=\int_{-1}^{\infty}e^{\alpha(1+\SSS)}(1+\SSS)^{-2+\alpha}\dd \SSS = e^{\alpha}\alpha^{1-\alpha}\int _{0}^{\infty}e^{-z}z^{\alpha - 2}\dd z = e^{-\alpha}\alpha^{1-\alpha}\Gamma(\alpha - 1).
\end{align}
Finally, $f/C$ gives Eq.~\eqref{eq: gamma pdf}. This process is repeated for every probability distribution $\{ f_k \}_{k=1}^5$, although the general formula is derived below.

\subsection{The general case}

A general one-dimensional SDE reads as:
\begin{equation}\label{eq:1}
    \dd x = V(x)\dd t + \sigma(x)\dd W_t,
\end{equation}
where $W_t$ is a standard Wiener process and $V$ and $\sigma \neq 0$ have the regularity so that solutions distribute according to a smooth probability density function \cite{gardiner2009,pavliotisbook2014}. The associated Fokker-Planck equation is:
\begin{equation}
\partial_t f = \partial_x \left[ - V(x)f + \frac{1}{2}\partial_x\left( \sigma^2(x)f \right) \right],
\end{equation}
where $f$ is a probability density and a function of $x$ and $t$. The stationary distribution of Eq.~\eqref{eq:1} is obtained by setting $\partial_tf = 0$, and solving for $f$:
\begin{align}
0 = \partial_x \left[ - V(x)f + \frac{1}{2}\partial_x\left( \sigma^2(x)f \right) \right].
\end{align}
Integrating from $ \ell $, where $f$ is assumed to vanish, to $x$ and rearranging to make the equation homogeneous,
\begin{equation}
\frac{\partial_xf(x)}{f(x)} = \frac{2V(x)}{\sigma^2(x)} - \frac{\partial_x\left(\sigma^2(x)\right)}{\sigma^2(x)}
\end{equation}
solving the indefinite integral on both sides yields:
\begin{equation}
    \log f(x) = \int^x \frac{2V(x)}{\sigma^2(x)} \dd x - \log\left(\sigma^2(x) \right),
\end{equation}
where ``$\int^x$" denotes the indefinite integral. We now take the exponentials on both sides:
\begin{equation}
f(x) = e^{\int^x \frac{2V(x)}{\sigma^2(x)}}\dd x\sigma^{-2}(x).
\end{equation}
One is left with finding the normalising constant.

\section{The square-root process}\label{app:square_root}

The condensational growth equation is easily integrated into the supersaturation equation, although its expression depends on the square-root of the initial particle size plus its fluctuating part due to supersaturation variations; see Eq.~\eqref{eq:int_diff}. If such fluctuations, here denoted as $\delta_{\SSS}(t)$, decorrelate instantly and have zero mean, such expression is rewritten as:
\begin{equation}
    r(t) = \sqrt{r^2(t)} = \sqrt{r(0)^2 + 2cA_{\circ}\int_0^t\SSS(s)\dd s} \approx \sqrt{r(0)^2 +\delta_{\SSS}(t)},
\end{equation}
where the variance of $\delta_{\SSS}(t)$, $\sigma^2_{r^2}$, is proportional to the diffusional constant $A_{\circ}$. Deriving the statistics of a squared-root stochastic process is difficult, although in the limit of variance fluctuations of $r^2$ begin small, or when the diffusional constant $A_{\circ}$ is small, we can derive the steady-state mean and variance of the square-root process as an expansion:
\begin{subequations}\label{eq:expansion_square_root}
\begin{align}
\mathbb{E}\left[ r \right] &= \mathbb{E}\left[ \sqrt{r^2} \right] = r(0) - \frac{1}{8}r(0)^{-3}\sigma^2_{r^2} + \mathcal{O}\left(\sigma_{r^2}^4\right)  \\
\mathrm{Var}\left[ r \right] &= \mathrm{Var}\left[ \sqrt{r^2} \right] = \mathbb{E}\left[ 
r^2 \right] - \mathbb{E}\left[ \sqrt{r^2} \right]^2 = \frac{1}{4r(0)^2}\sigma^2_{r^2} + \mathcal{O}\left( \sigma^4_{r^2} \right).
\end{align}
\end{subequations}
Surprisingly, the variance of the square-root process only scales inversely proportionally to $r(0)^2$.  This approach assumes that $r(0)$ is the mean radius and that fluctuations $\delta_{\SSS}(t)$ are so small that $r^2$ remains positive. This result is general and can be applied to any square-root random variable, with positive mean, and in the limit of small variance.

\begin{figure}[h]
    \centering
    \includegraphics[width=0.6\linewidth]{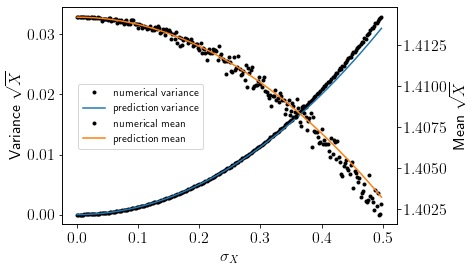}
    \caption{\label{fig:variance_square_root}\textbf{Variance and mean of the square-root random variable.} The black dots are calculated as follows: for each value of $\sigma_X$, the variance and mean of the random variable $\sqrt{X}$ are numerically estimated by taking the square root of $10^5$ draws of a normal random variable $X$ with mean $2$ and standard-deviation $\sigma_X$. The blue and orange solid lines indicate the truncated predictions of Eq.~\eqref{eq:expansion_square_root} for the variance and mean, respectively.}
\end{figure}

To support this analytical expansion, we numerically sampled an adimensional random variable $X$, normally distributed with mean $2$ and standard-deviation $\sigma_{X}$, where the latter takes 250 equispaced values between $10^{-4}$ and $0.5$. The sample is of size $10^5$ draws. With this set of data, we are able to numerically estimate the mean and variance of the square root random variable, $\sqrt{X}$, discarding all samples that gave negative values. The aim is to predict the variance and mean of $\sqrt{X}$, using the truncated expansions of Eq.~\eqref{eq:expansion_square_root} and the prescribed values of $\sigma_{X}$. The truncation is done at $\mathcal{O}\left( \sigma^4_{X} \right)$. Indeed, in Figure~\ref{fig:variance_square_root} the analytical predictions--- plotted in solid coloured curves--- match to a high degree of accuracy the numerically sampled random variable $\sqrt{X}$--- plotted in black dots---. When $\sigma_{X}$ becomes larger, a moderate deviation is observed, as expected.

\end{appendices}

\pagebreak

\pagebreak
{\small
\bibliographystyle{abbrv}
\bibliography{motr_project.bib}

\begin{thebibliography}{10}

\bibitem{arakawa1974}
A.~Arakawa and W.~H. Schubert.
\newblock Interaction of a cumulus cloud ensemble with the large-scale
  environment, part i.
\newblock {\em Journal of Atmospheric Sciences}, 31(3):674 -- 701, 1974.

\bibitem{bartlett1972}
J.~T. Bartlett and P.~R. Jonas.
\newblock On the dispersion of the sizes of droplets growing by condensation in
  turbulent clouds.
\newblock {\em Quarterly Journal of the Royal Meteorological Society},
  98(415):150--164, 1972.

\bibitem{bergeron_1935}
T.~Bergeron.
\newblock {\em On the physics of clouds and precipitation}, pages 156--–178.
\newblock Proces Verbaux de l'Association de Meteorologie. International Union
  of Geodesy and Geophysics, 1935.

\bibitem{butler_2007}
R.~W. Butler.
\newblock {\em Exponential families and tilted distributions}, page 145–182.
\newblock Cambridge Series in Statistical and Probabilistic Mathematics.
  Cambridge University Press, 2007.

\bibitem{Devenish2016}
B.~J. Devenish, K.~Furtado, and D.~J. Thomson.
\newblock {Analytical solutions of the supersaturation equation for a warm
  cloud}.
\newblock {\em Journal of the Atmospheric Sciences}, 73(9):3453--3465, 2016.

\bibitem{eytan2022}
E.~Eytan, A.~Khain, M.~Pinsky, O.~Altaratz, J.~Shpund, and I.~Koren.
\newblock Shallow cumulus properties as captured by adiabatic fraction in
  high-resolution les simulations.
\newblock {\em Journal of the Atmospheric Sciences}, 79(2):409 -- 428, 2022.

\bibitem{Field2014}
P.~R. Field, A.~A. Hill, K.~Furtado, and A.~Korolev.
\newblock {Mixed-phase clouds in a turbulent environment. Part 2: Analytic
  treatment}.
\newblock {\em Quarterly Journal of the Royal Meteorological Society},
  140(680):870--880, 2014.

\bibitem{gardiner2009}
C.~Gardiner.
\newblock {\em {Stochastic Methods: A Handbook for the Natural and Social
  Sciences}}.
\newblock Springer-Verlag Berlin, Heildelberg, 2009.

\bibitem{Hill2014}
A.~A. Hill, P.~R. Field, K.~Furtado, A.~Korolev, and B.~J. Shipway.
\newblock {Mixed-phase clouds in a turbulent environment. Part 1: Large-eddy
  simulation experiments}.
\newblock {\em Quarterly Journal of the Royal Meteorological Society},
  140(680):855--869, 2014.

\bibitem{khain_pinsky_2018}
A.~P. Khain and M.~Pinsky.
\newblock {\em Physical Processes in Clouds and Cloud Modeling}.
\newblock Cambridge University Press, 2018.

\bibitem{khvorostyanov_1999}
V.~I. Khvorostyanov and J.~A. Curry.
\newblock Toward the theory of stochastic condensation in clouds. part i: A
  general kinetic equation.
\newblock {\em Journal of the Atmospheric Sciences}, 56(23):3985 -- 3996, 1999.

\bibitem{Korolev2008}
A.~Korolev and P.~R. Field.
\newblock {The effect of dynamics on mixed-phase clouds: Theoretical
  considerations}.
\newblock {\em Journal of the Atmospheric Sciences}, 65(1):66--86, 2008.

\bibitem{Korolev2003}
A.~V. Korolev and I.~P. Mazin.
\newblock {Supersaturation of water vapor in clouds}.
\newblock {\em Journal of the Atmospheric Sciences}, 60(24):2957--2974, 2003.

\bibitem{pavliotisbook2014}
G.~A. Pavliotis.
\newblock {\em {Stochastic Processes and Applications}}, volume~60.
\newblock Springer, New York, 2014.

\bibitem{prabha_2011}
T.~V. Prabha, A.~Khain, R.~S. Maheshkumar, G.~Pandithurai, J.~R. Kulkarni,
  M.~Konwar, and B.~N. Goswami.
\newblock Microphysics of premonsoon and monsoon clouds as seen from in situ
  measurements during the cloud aerosol interaction and precipitation
  enhancement experiment (caipeex).
\newblock {\em Journal of the Atmospheric Sciences}, 68(9):1882 -- 1901, 2011.

\bibitem{prabhakaran_2020}
P.~Prabhakaran, A.~S.~M. Shawon, G.~Kinney, S.~Thomas, W.~Cantrell, and R.~A.
  Shaw.
\newblock The role of turbulent fluctuations in aerosol activation and cloud
  formation.
\newblock {\em Proceedings of the National Academy of Sciences},
  117(29):16831--16838, 2020.

\bibitem{risken}
H.~Risken.
\newblock {\em {The Fokker-Planck Equation}}.
\newblock Springer, second edition, 1989.

\bibitem{rodean1996}
H.~C. Rodean.
\newblock {\em Stochastic Lagrangian models of turbulent diffusion}.
\newblock Meteorological Monographs ; 48. American Meteorological Society,
  Boston, Massachusetts, 1st ed. 1996. edition, 1996.

\bibitem{rogersyau}
R.~R. Rogers and M.~K. Yau.
\newblock {\em {A Short Course in Coud Physics}}.
\newblock Pergamon Press, third edition, 1989.

\bibitem{sardina_2015}
G.~Sardina, F.~Picano, L.~Brandt, and R.~Caballero.
\newblock Continuous growth of droplet size variance due to condensation in
  turbulent clouds.
\newblock {\em Physical review letters}, 115(18):184501, 2015.

\bibitem{shaw_2003}
R.~A. Shaw.
\newblock Particle-turbulence interactions in atmospheric clouds.
\newblock {\em Annual Review of Fluid Mechanics}, 35(1):183--227, 2003.

\bibitem{siebert2017}
H.~Siebert and R.~A. Shaw.
\newblock Supersaturation fluctuations during the early stage of cumulus
  formation.
\newblock {\em Journal of the Atmospheric Sciences}, 74(4):975 -- 988, 2017.

\bibitem{siewert_bec_krstulovic_2017}
C.~Siewert, J.~Bec, and G.~Krstulovic.
\newblock Statistical steady state in turbulent droplet condensation.
\newblock {\em Journal of Fluid Mechanics}, 810:254–280, 2017.

\bibitem{squires_1952}
P.~Squires.
\newblock The growth of cloud drops by condensation.
\newblock {\em Australian Journal of Chemistry}, 1952.

\bibitem{ornstein_1930}
G.~E. Uhlenbeck and L.~S. Ornstein.
\newblock On the theory of the brownian motion.
\newblock {\em Phys. Rev.}, 36:823--841, Sep 1930.

\bibitem{zelinka2020}
M.~D. Zelinka, T.~A. Myers, D.~T. McCoy, S.~Po-Chedley, P.~M. Caldwell,
  P.~Ceppi, S.~A. Klein, and K.~E. Taylor.
\newblock Causes of higher climate sensitivity in cmip6 models.
\newblock {\em Geophysical Research Letters}, 47(1):e2019GL085782, 2020.

\end{thebibliography}
\nocite{*}
}
\end{document}